\pdfoutput=1
\documentclass[%
 reprint,
superscriptaddress,
 amsmath,amssymb,
 aps,
prb,
floatfix,
]{revtex4-2}

\usepackage{graphicx}
\usepackage{dcolumn}
\usepackage{bm}
\usepackage{pdfpages}
\usepackage{textcomp}

\makeatletter
\AtBeginDocument{\let\LS@rot\@undefined}
\makeatother

\begin{document}

\preprint{revtex4-2}

\title{Additional excitonic features and momentum-dark states in ReS\textsubscript{2}}

\author{A. Dhara}
\author{D. Chakrabarty}
\author{P. Das}
\author{A. K. Pattanayak}
\author{S. Paul}
\author{S. Mukherjee}
\author{S. Dhara}
 \email{sajaldhara@phy.iitkgp.ac.in}
\affiliation{
 Department of Physics, IIT Kharagpur, Kharagpur, WB 721302, India
}
\date{\today}

\begin{abstract}
	Unidirectional in-plane structural anisotropy in Rhenium-based dichalcogenides introduces a new class of 2-D materials, exhibiting anisotropic optical properties. In this work, we perform temperature dependent, polarization-resolved photoluminescence and reflectance measurements on several-layer ReS\textsubscript{2}. We discover two additional excitonic resonances (X\textsubscript{3} and X\textsubscript{4}), which can be attributed to splitting of spin degenerate states. Strong in-plane oscillator strength of exciton species X\textsubscript{1} and X\textsubscript{2} are accompanied by weaker counterparts X\textsubscript{3} and X\textsubscript{4} with similar polarization orientations. The in-plane anisotropic dielectric function has been obtained for ReS\textsubscript{2} which is essential for engineering light matter coupling for polarization sensitive optoelectronic devices. Furthermore, our temperature dependent study revealed the existence of low-lying momentum-forbidden dark states causing an anomalous PL intensity variation at 30 K, which has been elucidated using a rate equation model involving phonon scattering from these states. Our findings of the additional excitonic features and the momentum-dark states can shed light on the true nature of the electronic band structure of ReS\textsubscript{2}. 
\end{abstract}

\maketitle

\section{\label{sec:Introduction}Introduction}

In-plane structural anisotropy in individual layers of van der Waals materials produces electronic bandstructures that are unique in contrast to the family of Transition Metal Dichalcogenides (TMDCs) with in-plane rotational symmetry. Rhenium (Re) based  Group VII TMDCs like ReS\textsubscript{2} have garnered considerable attention because of their intriguing anisotropic optical, vibrational and electronic properties arising from reduced crystal symmetry \cite{aslanLinearlyPolarizedExcitons2016a,tongayMonolayerBehaviourBulk2014, aroraHighlyAnisotropicInPlane2017a,echeverryTheoreticalInvestigationsAnisotropic2018,jadczakExcitonBindingEnergy2019b,urbanNonEquilibriumAnisotropic2018,simUltrafastQuantumBeats2018a,olivaPressureDependenceDirect2019,liuIntegratedDigitalInverters2015a,cuiTransientAbsorptionMeasurements2015,zhouStackingOrderDriven2020,wangDirectIndirectExciton2019,simSelectivelyTunableOptical2016,hafeezRheniumDichalcogenidesReX2017}. This originates from ReS\textsubscript{2}’s distorted 1T structure \cite{wilsonTransitionMetalDichalcogenides1969}, where the extra electron from the Re atom  contributes to the strong Re-Re metal bond, forming  a zigzag chain along the b-axis. This results in higher electron mobility along the b-axis \cite{liuIntegratedDigitalInverters2015a}, and optical anisotropy manifesting itself as highly polarized photoluminescence (PL) and absorption due to two strongly bound exciton species X\textsubscript{1} and X\textsubscript{2}  with dipole moments along different in-plane directions \cite{aslanLinearlyPolarizedExcitons2016a,jadczakExcitonBindingEnergy2019b,urbanNonEquilibriumAnisotropic2018,simUltrafastQuantumBeats2018a}. These properties open the door to a class of polarization-sensitive, on-chip devices like  polarization controlled all-optical switches \cite{simLightPolarizationControlledConversion2019}, polarized LEDs \cite{wangPolarizedLightEmitting2020}, photodetectors \cite{zhangReS2BasedFieldEffectTransistors2015} and polarization-based quantum logic gates \cite{kwonAll2DReS2Transistors2019}. Additionally, for Group VI TMDCs the electronic and optical properties are strongly dependent on the number of layers of the crystal, and show a drastic change in the monolayer limit \cite{splendianiEmergingPhotoluminescenceMonolayer2010}. Conversely, ReS\textsubscript{2} with its weak interlayer coupling shows no such drastic change when going from bulk to monolayer \cite{tongayMonolayerBehaviourBulk2014}, making it ideal for multilayer photonic device applications. In order to create highly sensitive devices, however, a meticulous study of its anisotropic dielectric properties is required. Furthermore, there has been considerable debate about the nature of ReS\textsubscript{2}’s bandgap, with the recent consensus being that it is marginally indirect except in bilayer form \cite{urbanNonEquilibriumAnisotropic2018,olivaPressureDependenceDirect2019,wangDirectIndirectExciton2019,biswasNarrowbandAnisotropicElectronic2017,webbElectronicBandStructure2017}. Experimental studies that can shed light into this matter are essential.

\begin{figure*}[t]
	\includegraphics{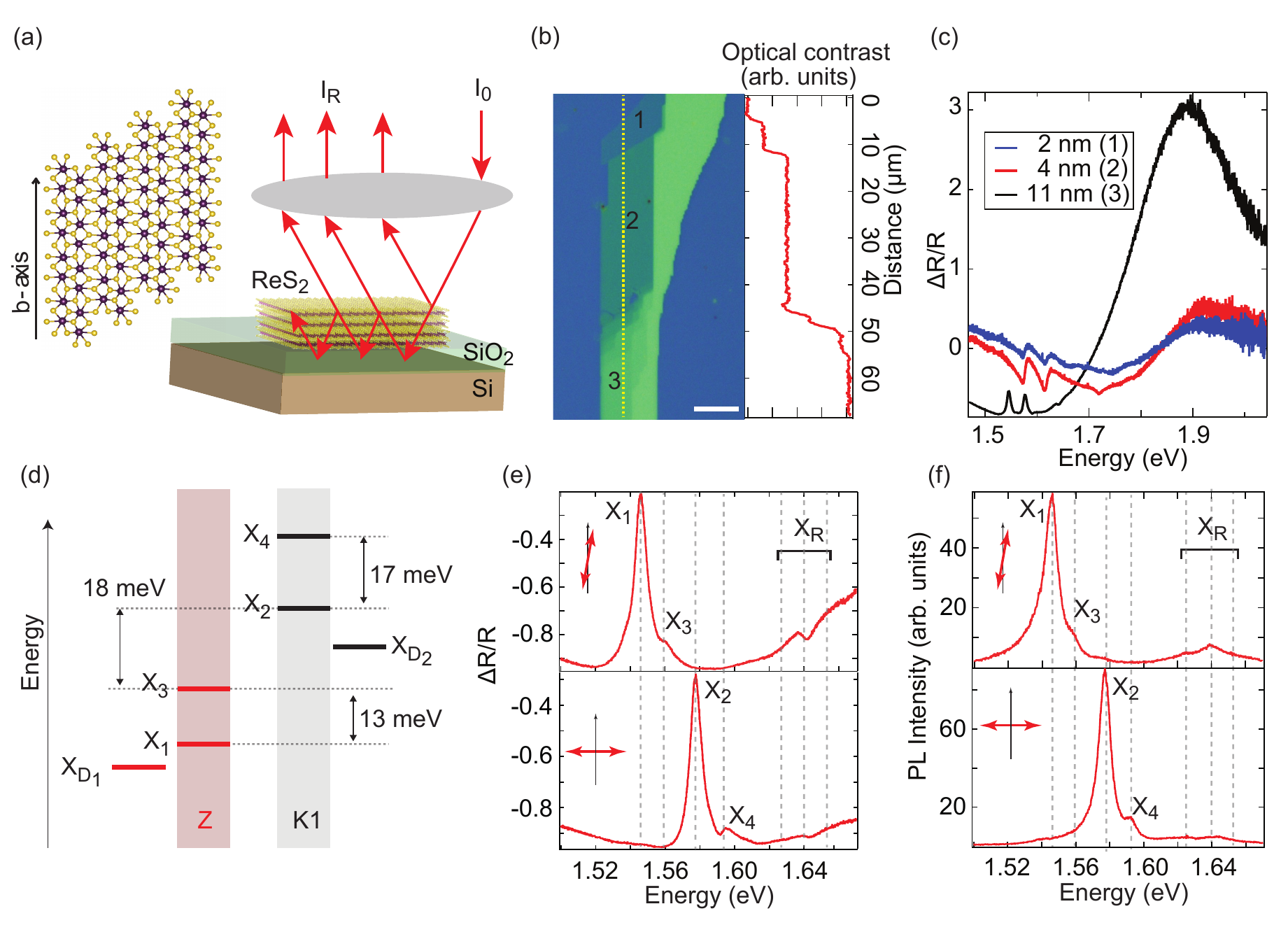}
	\caption{\label{fig:1}(a) Top and side view of the distorted 1T phase of monolayer  ReS\textsubscript{2}. The arrow indicates the b-axis of crystal orientation. b) Optical microscope image of exfoliated  ReS\textsubscript{2} transferred on Si/SiO\textsubscript{2} substrate. The white scale bar represents 10 $\mu$m. (left). Optical contrast along the dotted yellow line showing layer thickness (right). (c) Differential reflectance from unpolarized reflectance study at 3 different points in the sample marked as 1, 2 and 3 (2, 4 and 11 nm) in (b). Further studies are conducted on point 3, i.e., 11 nm ReS\textsubscript{2}. (d) Schematic showing the exciton complexes at Z and K1 points of the Brillouin Zone along with the low-lying momentum-dark states (e) Differential reflectance for two different incident polarizations w.r.t the b-axis as indicated in the inset of the top and bottom panel, showing four exciton peaks for 11 nm ReS\textsubscript{2}. (f) Polarized PL intensity measured at two different orientations of the analyzer w.r.t the b-axis as indicated in the inset of the top and bottom panel.  (e), (f) corresponds to the polarization state for which the contribution is minimized from X\textsubscript{2} and X\textsubscript{4} (top panel) and X\textsubscript{1} and X\textsubscript{3} (bottom panel) respectively. X\textsubscript{R} denotes the Rydberg excitations of X\textsubscript{1} and X\textsubscript{2}.}
\end{figure*}

In this work, we focus on high-resolution, polarization-resolved reflectance and PL at low temperature, discovering two additional exciton peaks that were predicted for ReS\textsubscript{2} \cite{echeverryTheoreticalInvestigationsAnisotropic2018} but not observed in previous studies. It is envisaged that these two shoulder peaks denoted by X\textsubscript{3} and X\textsubscript{4} are observed due to the splitting of spin degenerate excitonic states by combined effect of electron-hole exchange interaction, structural anisotropy, and spin-orbit coupling.  X\textsubscript{3} and X\textsubscript{4} appear on the higher energy sides in both the absorption and PL measurements with similar polarization orientation as X\textsubscript{1} and X\textsubscript{2} respectively. The anisotropic nature of ReS\textsubscript{2}’s dielectric properties is further demonstrated via polarization dependent reflectance. Transfer matrix method was utilized to understand the asymmetric excitonic lineshape in our reflectance from the ReS\textsubscript{2}/SiO\textsubscript{2}/Si dielectric stack as a function of layer thickness and thus extract the real and imaginary part of anisotropic refractive index of the material. In addition, we report excitation polarization dependence of the PL, which provides insight into the anisotropic absorption of this material. For excitation energy 1.88 eV, the integrated PL intensity from all four exciton species is maximized when excitation polarization is oriented along X\textsubscript{2}. Furthermore, our temperature-dependent study finds the four exciton peak positions are well resolved up to a temperature of 150 K, and their dipole orientations are preserved. Most intriguingly, we observe an anomalous PL intensity variation around 30 K, which necessitated the development of a model using rate equations, considering phonon scattering from low-lying dark exciton states. Our model provides evidence of the existence of indirect exciton states which are closely-lying below the bright excitons, confirming the quasi-indirect nature of ReS\textsubscript{2}’s bandgap. Our analysis can shed light on the understanding of the excitonic properties in this anisotropic material and provide useful information for photonics and optoelectronic device engineering with ReS\textsubscript{2}.

\section{\label{sec:Results}Results and Discussion}

\begin{figure*}[t]
	\includegraphics{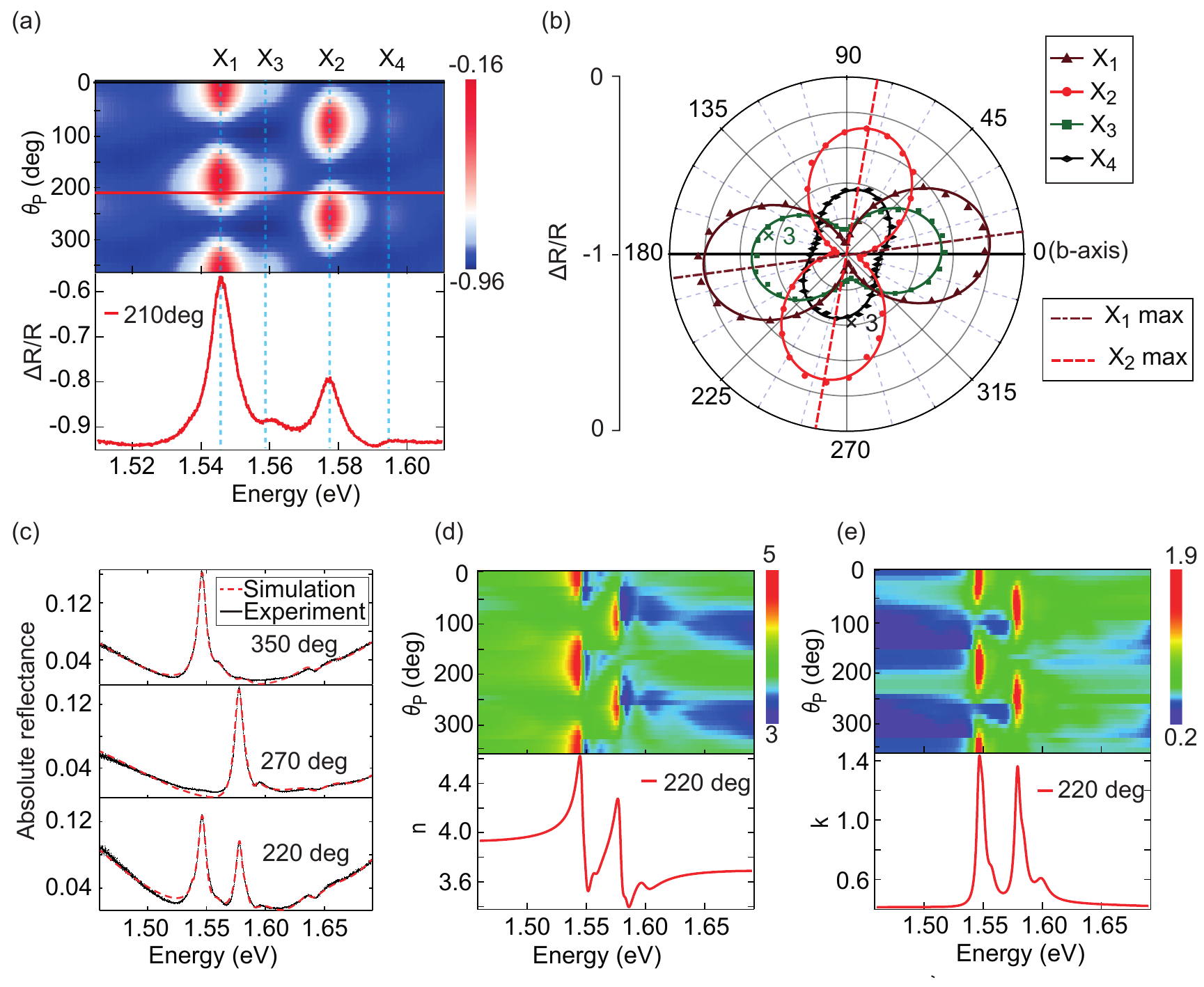}
	\caption{\label{fig:2}(a) Colour plot representing the differential reflectance spectrum as a function of incident polarization angle $\theta_{P}$ measured w.r.t. the b-axis for 11 nm ReS\textsubscript{2}. Line profile in the lower panel shows reflectance at a particular polarization angle where all four excitons can be observed. (b) Polar plots of differential reflectance for four different excitons X\textsubscript{1}, X\textsubscript{2}, X\textsubscript{3} and X\textsubscript{4} proportional to the oscillator strength (values for X\textsubscript{3} and X\textsubscript{4} are multiplied by 3 for clarity). (c) Polarization resolved absolute reflectance data (black) along with the simulation result (red) obtained from the transfer matrix method, shown for three different polarisation angles. (d), (e) Colour plots demonstrating the in-plane anisotropy in the real and imaginary part of the refractive index, obtained from the fitting of absolute reflectance at variable polarization. The lower panel shows the corresponding line profile obtained at particular angle of polarization.}
\end{figure*}

A schematic top view is shown in Fig. \ref{fig:1}(a) of a single layer of ReS\textsubscript{2}, where each molecular layer is a sandwich of a Re layer between two S layers. The direction of the b-axis is marked by a solid arrow which can be identified from the optical image during measurement. Sample was prepared by mechanical exfoliation technique, and dry transferred on to 340 nm SiO\textsubscript{2}/Si substrate. The optical image of the sample is shown in Fig. 1(b) where we can approximately identify the number of layers from the optical contrast along a line profile, which has been further verified using AFM (see Fig. S1 in the Supplemental Material \cite{SeeSupplementalMaterial}). On-chip crystallographic orientation can be estimated by observing the sample edges, utilizing the fact that ReS\textsubscript{2} is likely to cleave along the axis containing the covalent Re-S bonds, which is parallel to the b axis \cite{aslanLinearlyPolarizedExcitons2016a,liangOpticalAnisotropyAudoped2009a}.    

Reflectance measurement was performed using a broadband halogen source with a spot size of $\sim$3 \textmu m at three different points of the sample as shown in Fig. 1(b): Points 1, 2 and 3, being 2 nm, 4 nm and 11 nm thick respectively. At 11 nm thickness (point 3), ReS\textsubscript{2} approaches its bulk character, wherein PL intensity becomes independent of the number of layers \cite{aslanLinearlyPolarizedExcitons2016a,tongayMonolayerBehaviourBulk2014}. However, its 2D character is preserved since excitons are confined mostly in single layers \cite{aroraHighlyAnisotropicInPlane2017a}. 

\begin{figure*}[t]
	\includegraphics{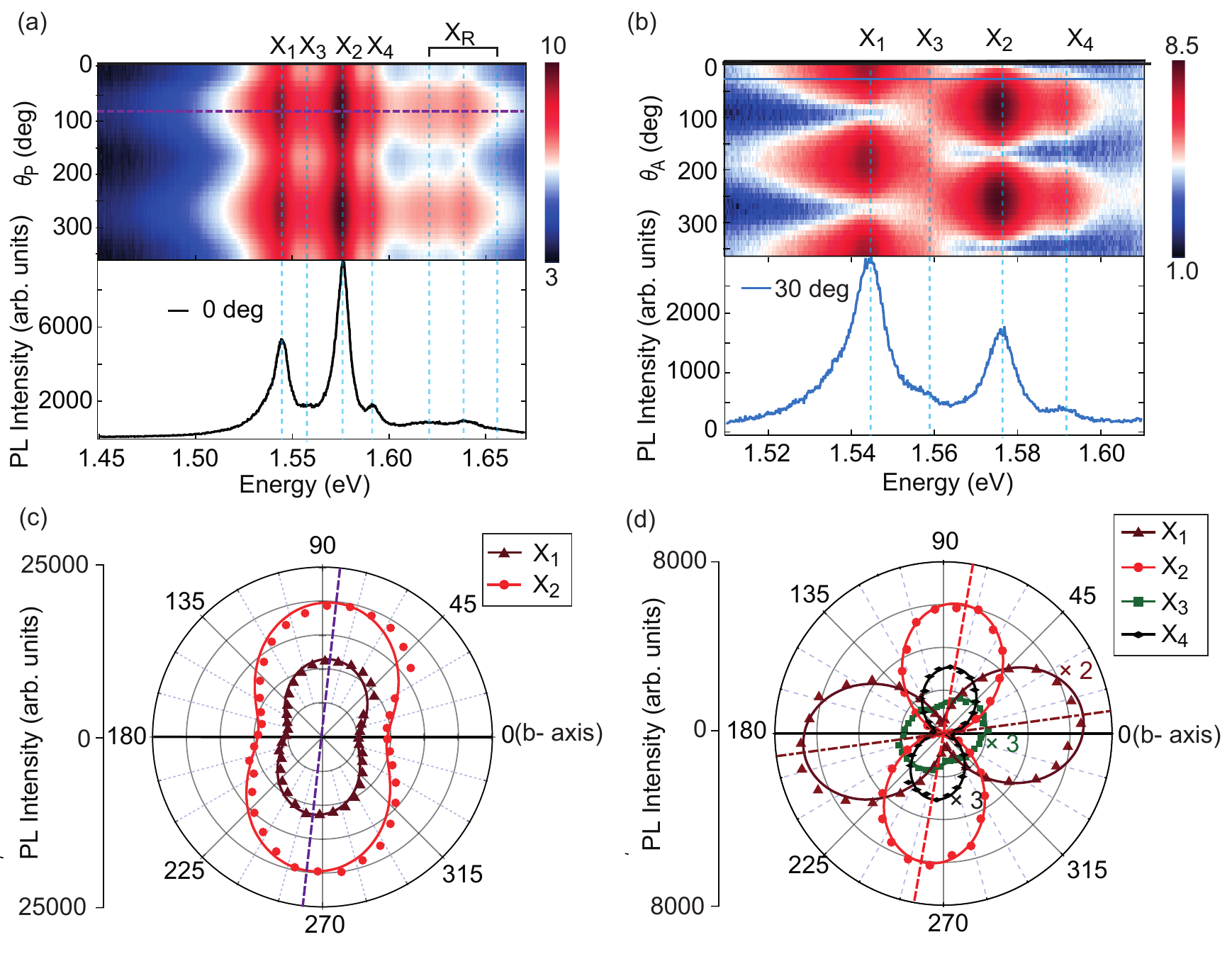}
	\caption{\label{fig:3}(a) Colour plot showing the PL spectrum in the log scale as a function of excitation laser polarization angle $\theta_{P}$ w.r.t. the b-axis for 11 nm ReS\textsubscript{2} when no analyzer is placed at the output path. The dotted line indicates the polarization angle at which the overall PL intensity is maximum. The linear scale PL spectrum when the excitation laser aligned with the b-axis is shown in the lower panel.  (b) Colour plot showing the PL spectrum in the log scale as a function of angle of the analyzer (w.r.t. b-axis) obtained at a fixed polarization of the excitation laser which maximize the PL intensity. For the same excitation polarization, a linear scale PL spectrum at 30$^\circ$ angle of the analyser is shown in the lower panel. (c) Polar plot representing the PL intensity variation for two most prominent peaks, X\textsubscript{1} and X\textsubscript{2} as a function of the excitation polarization angle. (d) Polar plot representing the PL intensity variation of X\textsubscript{1}, X\textsubscript{2}, X\textsubscript{3}, and X\textsubscript{4} as a function of the angle of the analyzer $\theta_{P}$ at fixed excitation polarization (Intensity of X\textsubscript{1}, X\textsubscript{3} and X\textsubscript{4} are multiplied by factors of 2, 3 and 3 respectively for clarity). The dotted lines indicate corresponding angle of maximum PL for X\textsubscript{1} and X\textsubscript{2} excitons.}
\end{figure*}

Excitonic features in PL and reflectivity are too broad at room temperature to resolve all the peaks, therefore, the sample was cooled down to 3.2 K in a closed cycle cryostat. The reflectivity measured at point 1 and 2 show an asymmetric lineshape about the excitonic resonance, as shown in Fig. 1(c). We see a pronounced red shift of exciton peak positions from few layers to bulk limit in agreement with previous reports \cite{aslanLinearlyPolarizedExcitons2016a,jadczakExcitonBindingEnergy2019b,urbanNonEquilibriumAnisotropic2018}. In reflectivity data, we plot the differential reflectance, that is, $\frac{\Delta R}{R}=(R_{ReS_2+SiO_2/Si}-R_{SiO2/Si})/R_{SiO2/Si}$, where $R_{ReS_2+SiO_2/Si}$ and $R_{SiO_2/Si}$ are the reflectance spectra from the sample and the SiO\textsubscript{2}/Si substrate respectively. Due to interference in the multilayer film system consisting of Si, SiO\textsubscript{2} and ReS\textsubscript{2} as shown in the schematic in Fig. 1(a), a broad anti-reflection dip coincides with the excitonic resonances. This makes the 11 nm ReS\textsubscript{2} (at point 3) ideal to probe well-resolved exciton peaks in reflectance, which almost mimics the PL spectrum. Chosen thickness of the SiO\textsubscript{2} layer and the dielectric properties of 11 nm ReS\textsubscript{2} make these peaks show a background free, almost symmetric Fano lineshape. In contrast, the reflectivity measured at point 1 and 2 show asymmetric Fano lineshape, which is a result of interference with the background reflectivity with a relative phase change of $\pi$ across the exciton resonance \cite{limonovFanoResonancesPhotonics2017}. The Fano asymmetry parameter varies across the three points. The peaks at higher energy range are the higher-order Rydberg series of the excitons \cite{aslanLinearlyPolarizedExcitons2016a,jadczakExcitonBindingEnergy2019b}. What follows below are based on experimental results conducted on the 11 nm ReS\textsubscript{2}. 

We observe four peaks at 1.545$\pm$0.001 (X\textsubscript{1}), 1.558$\pm$0.001 (X\textsubscript{3}), 1.576$\pm$0.001 (X\textsubscript{2}) and 1.593$\pm$0.001 (X\textsubscript{4}) eV. The peaks at higher energy range are the higher-order Rydberg series of the excitons, denoted by X\textsubscript{2} \cite{aslanLinearlyPolarizedExcitons2016a,jadczakExcitonBindingEnergy2019b}. The newly observed peaks X\textsubscript{3} and X\textsubscript{4} are in contrast to what has been observed earlier in ReSe2 \cite{aroraHighlyAnisotropicInPlane2017a}, since they appear at the higher energy side of X\textsubscript{1} and X\textsubscript{2} respectively.  However, from the resemblance between these additional peaks and the similar peaks for ReSe2, we speculate their origin is the splitting of singlet and triplet states of excitons due to electron-hole exchange interaction. Other plausible contributions for such splitting are the broken rotational symmetry due to structural anisotropy and spin-orbit coupling in ReS\textsubscript{2}.  As shown in Fig. 1(d), we attribute the higher energy prominent exciton transition X\textsubscript{2} to the K1 point, and X\textsubscript{1} to the Z point of the Brillouin Zone, using the results of a recent, comprehensive ab-initio calculation \cite{olivaPressureDependenceDirect2019}. X\textsubscript{D1} and X\textsubscript{D2} are low-lying momentum-dark states nearly degenerate with X\textsubscript{1} and X\textsubscript{2} respectively. They will be of interest when considering temperature dependence of PL intensity later in this paper. Figures 1(e) and 1(f) show the polarization resolved differential reflectance and PL at a polarization angle of -10$^{\circ}$  (top panels) and 90$^{\circ}$  (bottom panels) respectively, where all exciton peaks as seen in the reflectivity are also observed in PL with same energy positions. The 2 meV Stokes shift between PL emission and absorption peak (Fig. S2 in the Supplemental Material \cite{SeeSupplementalMaterial}) and the strong PL intensity indicates the pristine quality of our sample. We see no change in its optical properties over time and over multiple cooling cycles (Fig. S3 \cite{SeeSupplementalMaterial}).

In the polarization resolved reflectance measurement, linearly polarized white light was used as the source and the reflected light was collected via 0.7 NA objective lens, while the excitation polarization angle $\theta_{P}$ is varied with respect to (w.r.t.) the b-axis using a half-wave plate. The experimental result for 11 nm ReS\textsubscript{2} is shown in the colour plot in Fig. \ref{fig:2}(a). A line plot in the bottom panel is chosen at an angle where we see all the four exciton resonances. The polar plot Fig. 2(b) shows that the two exciton species X\textsubscript{1} and X\textsubscript{2} are polarized at angles of 6$^{\circ}$ and 79$^{\circ}$  respectively w.r.t the b-axis (For fitting function see Supplemental Material \cite{SeeSupplementalMaterial}), which agrees with earlier reported values within experimental error \cite{aslanLinearlyPolarizedExcitons2016a,olivaPressureDependenceDirect2019,hoCompleteseriesExcitonicDipole2019}. X\textsubscript{3} and X\textsubscript{4} follow the polarization variation of X\textsubscript{1} and X\textsubscript{2} respectively.

As shown in Fig. 1(a), we model the polarization resolved reflectance via transfer matrix technique \cite{blakeMakingGrapheneVisible2007,liMeasurementOpticalDielectric2014a,zhangMeasuringRefractiveIndex2015,hsuThicknessDependentRefractive2019,kuzmenkoKramersKronigConstrained2005}  (see Supplemental Material \cite{SeeSupplementalMaterial}) to obtain the reflectance spectrum of the ReS\textsubscript{2}/SiO\textsubscript{2}/Si stack and hence extract the frequency dependent dielectric function of ReS\textsubscript{2}. We fit the polarization-resolved reflectance to obtain the dielectric function $\epsilon(\omega)$ for every polarization. The dielectric function is given by $\epsilon(\omega)=\epsilon_b+\sum\frac{f_i}{\omega_{0i}^2-\omega^2-i\omega\gamma_i}$, where $\epsilon_b$, $f_i$, $\omega_{0i}$ and $\gamma_i$ are the background dielectric constant, oscillator strength, resonance frequency and the linewidth of the i\textsuperscript{th} oscillator, and the summation is over all exciton resonances.  Fig. 2(c) shows absolute reflectance spectra obtained at three different polarization angles of the linearly polarized incident beam, along with their theoretical fits. All the fitted parameters are given in Tables S1-S4 in Supplemental Material  \cite{SeeSupplementalMaterial}. 

The anisotropic refractive index $\tilde{n}\left(\hbar\omega\right)=n\left(\hbar\omega\right)+ik\left(\hbar\omega\right)$, where, $n$ and $k$ are the real and imaginary parts of refractive index, are shown in the colour plots Figs. 2(d) and 2(e) respectively. Interestingly, as shown in Fig. 2(c), we observed that the transfer matrix model best fit our experimental data with the polarization resolved reflectance only. However, in the Supplemental Material (Fig. S4 \cite{SeeSupplementalMaterial}) we show that it is not possible to obtain a good fit for unpolarized reflectance data with this model. As the unpolarized reflectance is a result of averaging over all polarization, it cannot be fitted by an effective value of the dielectric constant, since it is in truth highly anisotropic. 

A 660 nm laser with a spot size of $\sim$1 \textmu m is used to excite the sample with variable polarization w.r.t the b-axis. At first, the integrated PL was collected directly at the spectrometer slit without any analyzer at the output port. We discover that the overall PL intensity varies with the excitation polarization direction as shown in the colour plot in Fig. \ref{fig:3}(a). The intensity of the overall spectrum is modulated; however, the intensity ratio of X\textsubscript{1} and X\textsubscript{2} remains unchanged as the incident polarization is varied. Intensity from all exciton species is maximum at a particular polarization of excitation laser, corresponding to an in-plane direction which turns out to be the polarization direction of the X\textsubscript{2} exciton, within experimental error. This is a consequence of absorption at excitation energy 1.88 eV also being anisotropic, peaking at the angle along X\textsubscript{2}. This can be predicted from the angle dependent absorption obtained from our fitting (Fig. S5 \cite{SeeSupplementalMaterial}), and agrees with theoretical calculations \cite{echeverryTheoreticalInvestigationsAnisotropic2018}. The incident polarization dependence for two other lower excitation energies (1.76 and 1.70 eV) closer to X\textsubscript{2} resonance was also tested, which showed slightly different behavior (Fig. S6 \cite{SeeSupplementalMaterial}). The anisotropy becomes more pronounced as the excitation energy comes closer to the exciton resonances.

\begin{figure*}[t]
	\includegraphics{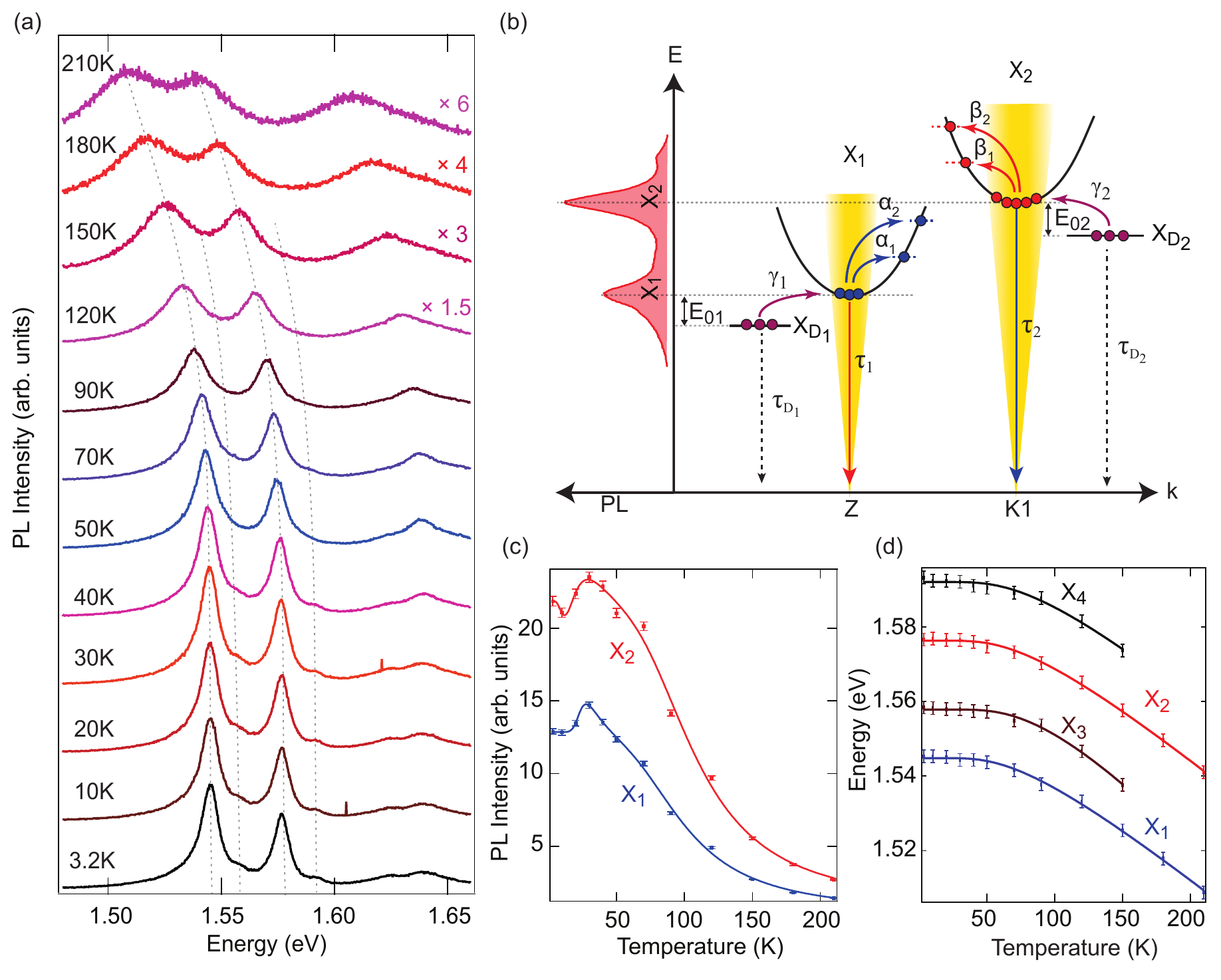}
	\caption{\label{fig:4}(a) PL spectra measured at different temperatures for a fixed angle of the analyser and excitation polarization. The dotted lines are a guide to eye. (b) Schematic diagram for the rate equation model. Curved arrows indicate all the phonon scattering processes considered in the model. Black dotted lines indicate non-radiative decay channel. Yellow region indicates the light cone for momentum allowed optical transitions. The solid straight arrows represent radiative decay channels. (c) Variation of PL intensity with temperature showing a local maxima around 30 K. Blue and red solid lines are fits obtained from the rate equation model. (d) Peak positions for the four excitons as a function of temperature. Solid lines are the fitted curves from the theoretical model.}
\end{figure*}

Next, we keep the excitation polarization fixed at an angle for which the overall PL is maximum and measure the polarization state of the PL via an analyzer placed before the spectrometer slit. Data shown in Fig. 3(b) was recorded while rotating the analyzer angle $\theta_{A}$ from 0$^{\circ}$  to {360}$^{\circ}$, where zero is along the b-axis. From the polar plot in Fig. 3(d), we find the two dominant emission peaks at X\textsubscript{1} and X\textsubscript{2} are polarized w.r.t the b-axis at angles of 6$^{\circ}$ and {82}$^{\circ}$ respectively, which is same as the angles obtained from reflectance considering experimental error. It is observed as shown in Figs. 2(b) and 3(d) that the two shoulder peaks at X\textsubscript{3} and X\textsubscript{4} follow the same polarization directions as X\textsubscript{1} and X\textsubscript{2} respectively.  

We perform a temperature-dependent PL measurement, the results of which are plotted in Fig. \ref{fig:4}(a). The analyzer is fixed throughout at an angle for which all four exciton peaks are visible. The two shoulder peaks X\textsubscript{3} and X\textsubscript{4} are not resolvable above 150 K since they are dominated by X\textsubscript{1} and X\textsubscript{2} whose linewidths increase with temperature. We observe an anomalous temperature variation of PL intensity as shown in Fig. 4(c), with a maxima in intensity near 30K, which can be attributed to the presence of low-lying dark states (X\textsubscript{D1}, X\textsubscript{D2}). Spin-forbidden dark excitons  \cite{molasBrighteningDarkExcitons2017} have recently been reported to be the cause of anomalous  temperature dependent PL in Group VI TMDCs  \cite{aroraDarkTrionsGovern2020}. On the other hand, momentum-forbidden low-lying states have been hypothesized before to explain the ratio of X\textsubscript{1} and X\textsubscript{2} populations not following the expected Boltzmann distribution \cite{aslanLinearlyPolarizedExcitons2016a,urbanNonEquilibriumAnisotropic2018}. This is also evident in Fig. \ref{fig:3} where it is observed that the X\textsubscript{1} PL intensity is lower than X\textsubscript{2} which is persistent even at higher temperatures as shown in Fig. 4(c).  Time-resolved measurements have also revealed that the radiative lifetimes of excitons to be less than 10 ps \cite{wangDirectIndirectExciton2019}, indicating the excitons are not thermalized. 

To gain a comprehensive understanding of the temperature variation of this hot photoluminescence, we propose a rate equation model involving the bright (X\textsubscript{1}, X\textsubscript{2}) and dark states (X\textsubscript{D1}, X\textsubscript{D2}) as shown in the schematic diagram Fig. 4(b), where excitons in these states are generated via continuous wave (CW) pumping. Two kinds of phonon scattering are important in this model which are discussed below. First, scattering of excitons from X\textsubscript{D1} to X\textsubscript{1} and X\textsubscript{D2} to X\textsubscript{2} with rates $\gamma_1$ and $\gamma_2$ respectively, where $E_{01}$ and $E_{02}$ are the energy of the phonons involved. Second, $\alpha_1$, $\alpha_2$ and $\beta_1$, $\beta_2$ are the scattering rates from X\textsubscript{1} and X\textsubscript{2} to momentum-forbidden dark states outside the light cone. Excitons scatter to a continuum of states outside the light cone via phonons of all available energies. To simplify the model, we have considered only two different phonon energies per exciton state - $E_{1n}$, $E_{1m}$ for X\textsubscript{1} and $E_{2n}$, $E_{2m}$ for X\textsubscript{2}. This effectively accounts for the scattering at both low temperature and high temperature regimes (see Fig. S10 in the Supplemental Material \cite{SeeSupplementalMaterial}).

We assume $\tau_1$ and $\tau_2$ are the radiative recombination times from X\textsubscript{1} and X\textsubscript{2}, and $\tau_{D1}$ and $\tau_{D2}$  are the non-radiative recombination times from X\textsubscript{D1} and X\textsubscript{D2}. By solving the rate equation under CW excitation (see Supplemental Material \cite{SeeSupplementalMaterial}), the PL intensity for X\textsubscript{1} and X\textsubscript{2} is obtained as a function of temperature, and fitted with the experimental data as shown in Fig. 4(c). From this model we find the two momentum-dark states X\textsubscript{D1} and X\textsubscript{D2} are present 16 meV and 5 meV below X\textsubscript{1} and X\textsubscript{2} states respectively. These dark states strongly indicate a quasi-indirect band gap at Z and K1 point in the Brillouin zone. This is in agreement with ab-initio calculations which indicate ReS\textsubscript{2} is marginally indirect \cite{urbanNonEquilibriumAnisotropic2018,olivaPressureDependenceDirect2019}.

When temperature is initially increased (3.2 K$<$T$<$30 K) X\textsubscript{1} and X\textsubscript{2} are thermally populated from X\textsubscript{D1} and X\textsubscript{D2} states by absorbing phonons. At the same time, X\textsubscript{1} and X\textsubscript{2} excitons are scattered from the radiative window to outside the light cone, where they recombine non-radiatively, by absorbing phonons of energy 17 meV and 3.2 meV respectively. In this temperature regime, we observe dark states are appreciably more populated than bright states on laser excitation, which results in a net gain in excitons scattered to the bright states. This causes PL intensity to increase to a maximum at 30K. At higher temperatures, the scattering process is mainly dominated by phonons of energies 52 meV and 41 meV from X\textsubscript{1} and X\textsubscript{2} states respectively. Scattering to dark states increases more rapidly than that to bright states after 30 K, causing PL intensity to decrease. All phonon energies obtained, except for 5 and 3.2 meV, have recently been reported for ReS\textsubscript{2} \cite{mccrearyIntricateResonantRaman2017a}. From our fitting we also notice that the effective scattering time from X\textsubscript{1} and X\textsubscript{2} is much more than their recombination times, even up to high temperatures, which corroborates the fact that X\textsubscript{1} and X\textsubscript{2} emit hot photoluminescence. 

We plot the observed redshift of exciton peak positions with temperature, and fit with the model describing the temperature dependence of a semiconductor bandgap \cite{odonnellTemperatureDependenceSemiconductor1991}: $E_g\left(T\right)=E_g\left(0\right)-S\hbar\omega\left[\coth{\left(\frac{\hbar\omega}{2kT}\right)}-1\right]$. Here $E_g\left(0\right)$ is the exciton resonance energy at T = 0 K, S is a dimensionless electron-phonon coupling constant, and $\hbar\omega$ is the average phonon energy. The fitted parameters are $E_g\left(0\right)$, $S$ and $\hbar\omega$. The values of $S$ are 1.74, 1.78, 2.42 and 1.83 for X\textsubscript{1}, X\textsubscript{2}, X\textsubscript{3} and X\textsubscript{4} respectively, and $\hbar\omega$ is around 20$\pm$3 meV, concurring with earlier reports \cite{hoAbsorptionedgeAnisotropyReS1998,hoOpticalStudyStructural2005}. Energy separation between the four exciton peaks remains constant with temperature as can be seen in Fig. 4(d). The angle between the excitons’ dipole moment orientation and b-axis remains unchanged with temperature indicates the temperature independent anisotropy (Fig. S7 \cite{SeeSupplementalMaterial}). Finally, temperature dependent reflectance measurements were used to acquire the corresponding dielectric function using the transfer matrix method (Fig. S8 \cite{SeeSupplementalMaterial}). $\epsilon_1$ shows significant change only near the exciton resonances.

\section{\label{sec:Conclusion}Conclusion}

In conclusion, we observed two additional exciton shoulder peaks at higher energy sides of X\textsubscript{1} and X\textsubscript{2} which are attributed to splitting of spin degenerate exciton states. The ab initio calculations required to systematically investigate the exact origin of this splitting is beyond the scope of this work, and may be carried out in future studies. The transfer matrix method is utilized to extract the in-plane anisotropic complex dielectric function of several-layer ReS\textsubscript{2} in unprecedented detail, which is crucial for modelling photonic devices or further experiments that make use of its anisotropic optical properties. Our temperature dependent study reveals an anomalous temperature variation in the PL intensity which underpins the existence of low-lying dark states, indicating quasi-indirect band gap in this system. The proposed mechanism considering the dynamical processes not only explains the observed temperature variation but also estimates the phonon energies supported by previous reports. It also establishes temperature dependent study as an accessible method to probe the existence of low-lying optically dark states.  

\begin{acknowledgments}
We acknowledge Dr. D. K. Goswami and his group for the AFM measurement on the sample. S. D. acknowledges SERB Ramanujan Fellowship, ISIRD grant, IIT Kharagpur and MHRD for the funding and support for this work. D. C. acknowledges CSIR for the financial assistance. We thank Dr. C. Chakraborty, and Dr. P. K. Chakraborty for their valuable comments on this work.    
\end{acknowledgments}


\providecommand{\noopsort}[1]{}\providecommand{\singleletter}[1]{#1}%



\section*{Supplementary material}

# Supplemental Material for

# Additional excitonic features and momentum-dark states in ReS₂

A. Dhara[1†], D. Chakrabarty[1†], P. Das[1], A. K. Pattanayak[1], S. Paul[1], S. Mukherjee[1], and S. Dhara[1*]

*¹Department of Physics, IIT Kharagpur, Kharagpur, WB 721302, India.*

**List of Contents:**

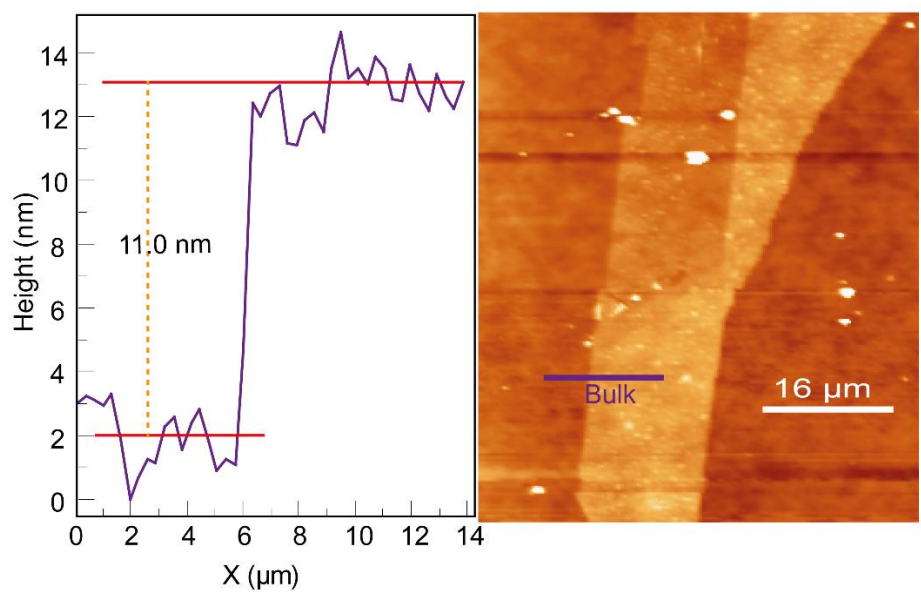

**FIG. S1.** (Right) Atomic force microscope image of ReS₂ sample. (Left) The thickness profile along the purple line in the AFM image. Thickness of the bulk is 11nm.

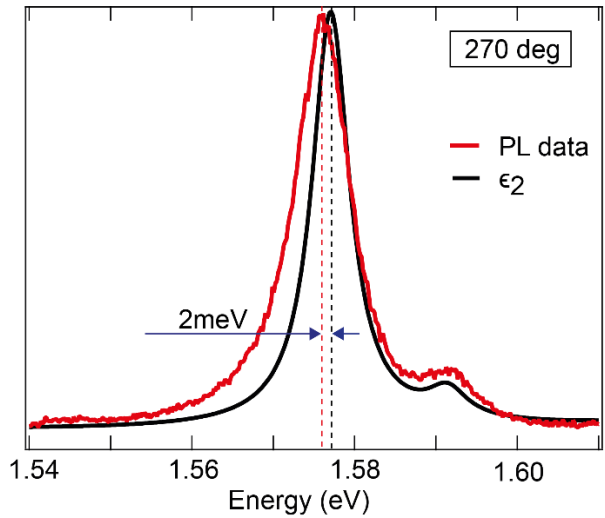

**FIG. S2.** The peak position shift (Stokes shift) between PL and absorption for $X_2$, found to be 2 meV. Red curve shows the PL for output polarization where the contribution from $X_1$ is minimum, and the black curve shows extracted absorption for incident white light for the same case.

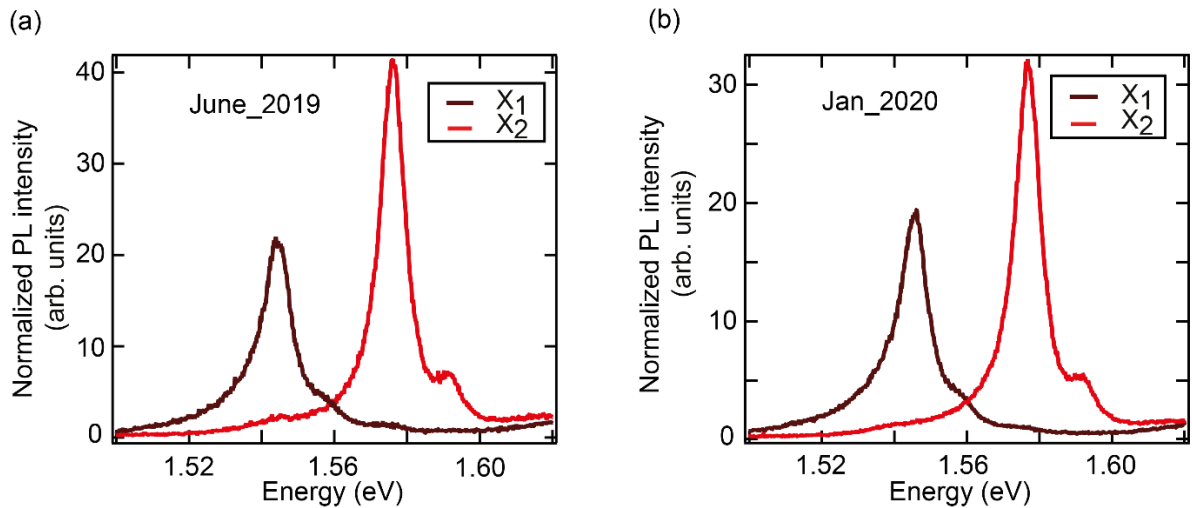

**FIG. S3.** The change in sample quality over time is represented in the figure in terms of change in PL response. (a) The brown and red colors are representing $X_1$ and $X_2$ PL spectrum respectively. (b) The measurement was taken 6 months after figure (a). The sample was kept in vacuum at room

temperature. We have seen there is a small change in PL intensity, but there is no defect of impurity-caused effects visible in the PL spectrum. This shows the robustness of ReS$_2$ as an optical material.

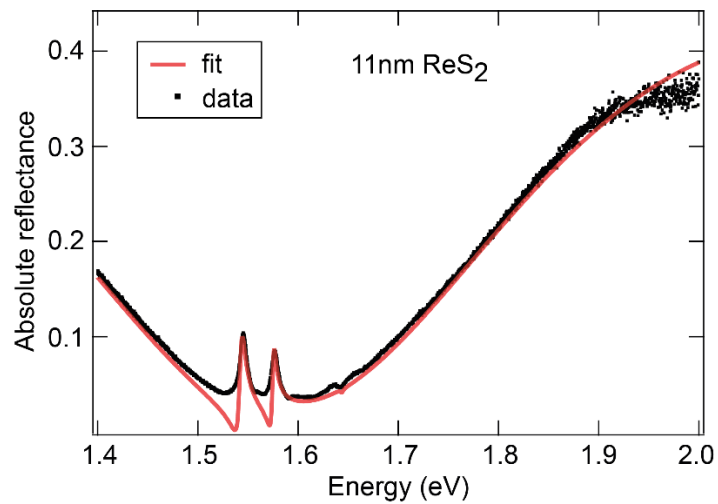

**FIG. S4.** Absolute reflectance of 11 nm ReS$_2$ on SiO$_2$/Si (344 nm) taken at 3.2K using unpolarized light, along with the fit (red line) obtained using transfer matrix method. A good fit cannot be obtained near the excitonic resonances using this method due to the anisotropic nature of the refractive index.

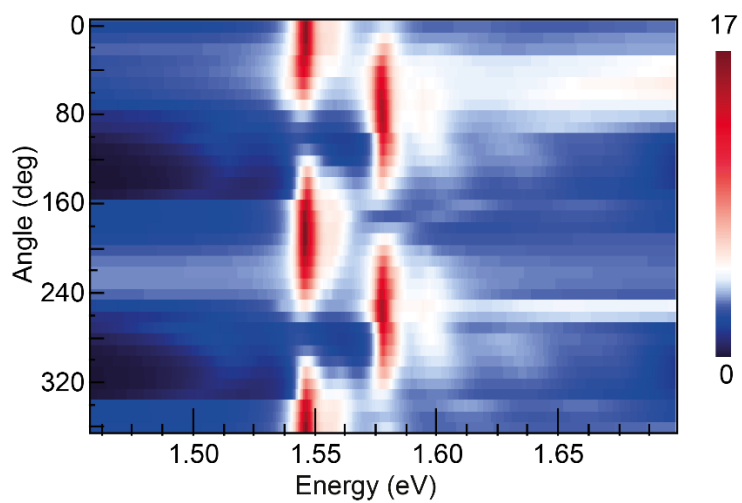

**FIG. S5.** Color plot showing complex part of the fitted dielectric function vs input polarization angle and reflected photon energy.

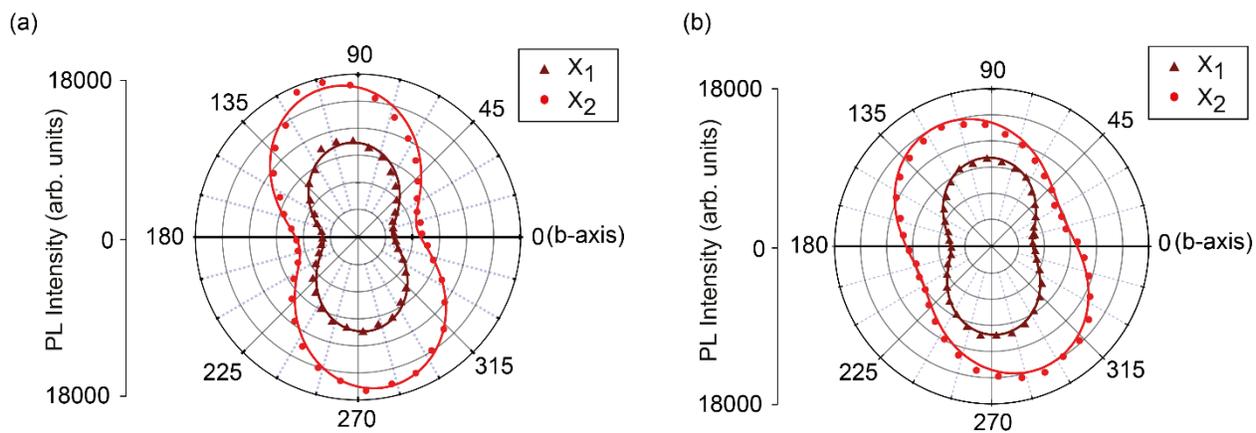

**FIG. S6.** Input polarization dependent PL measurement, for excitation energy (a) 1.76 eV and (b) 1.70 eV.

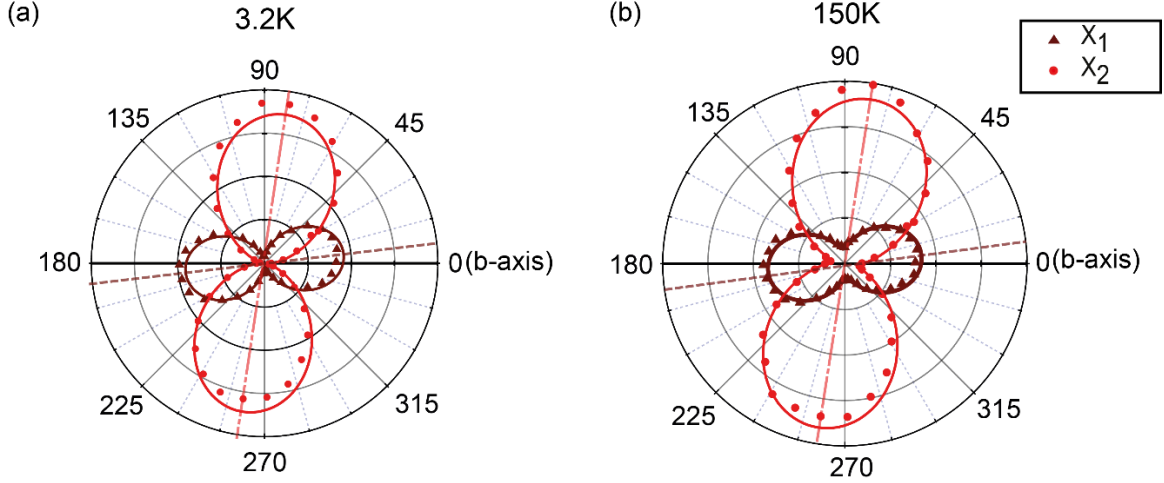

**FIG. S7.** Polarization dependent PL spectra for different temperatures. The PL intensity of $X_1$ and $X_2$ are fitted with the function given in Supplementary Note 2. (a) For 3.2K, $X_1$ is maximum

at $6.1996 \pm 0.476°$ and $X_2$ is maximum at $82.417 \pm 0.254°$. (b) For 150K the maximum values are $5.9268 \pm 0.68°$ and $80.475 \pm 0.563°$ respectively. The change in polarization angle in $X_1$ and $X_2$ with temperature is within the experimental error. While carrying out temperature dependent PL measurements, we have found that there is no significant variation of the orientations of $X_1$ and $X_2$, despite the red-shift of excitonic resonance energy.

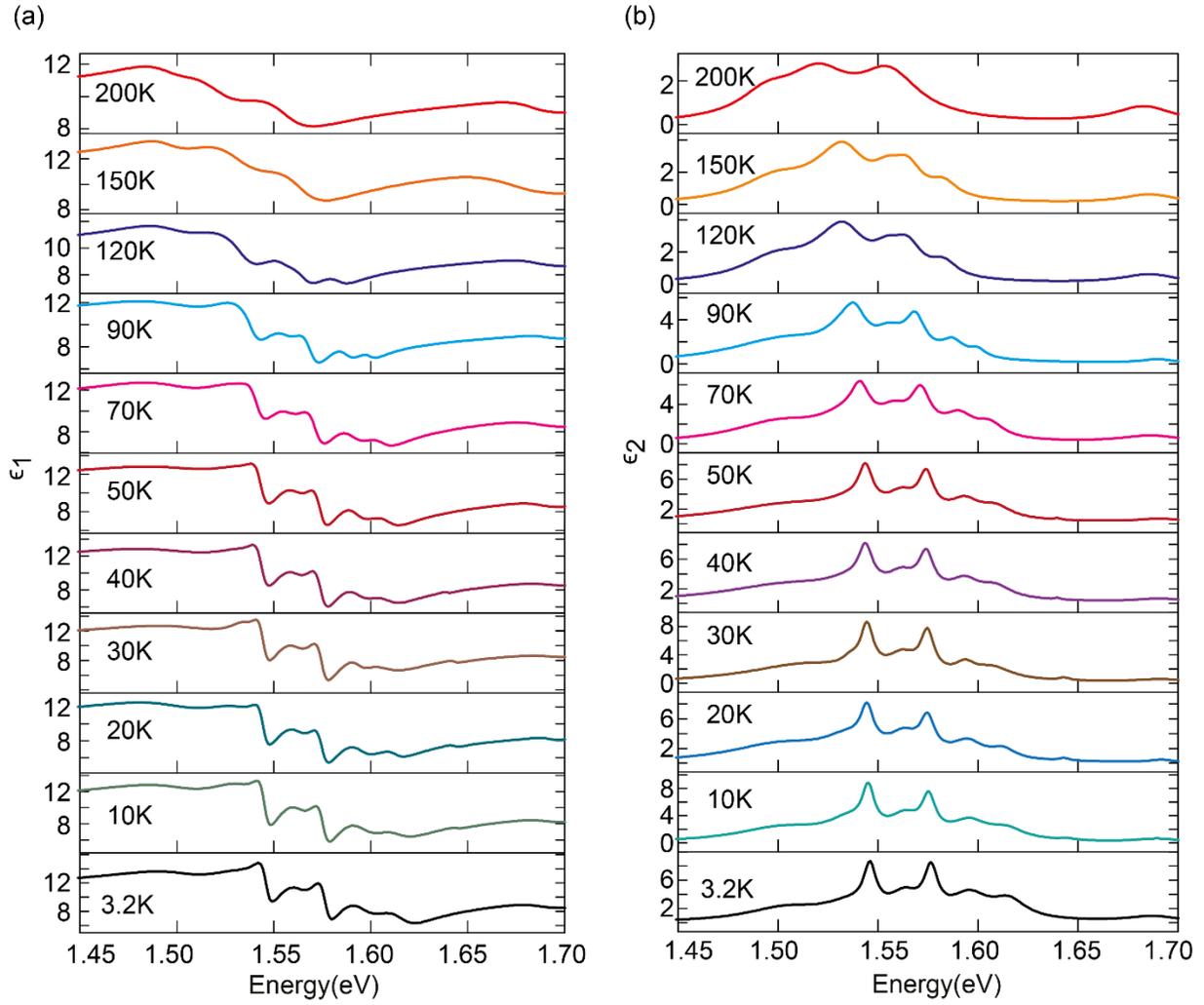

**FIG. S8**. Temperature dependent dielectric function for 11 nm ReS₂. (a) Real part and (b) Imaginary part of the dielectric function.

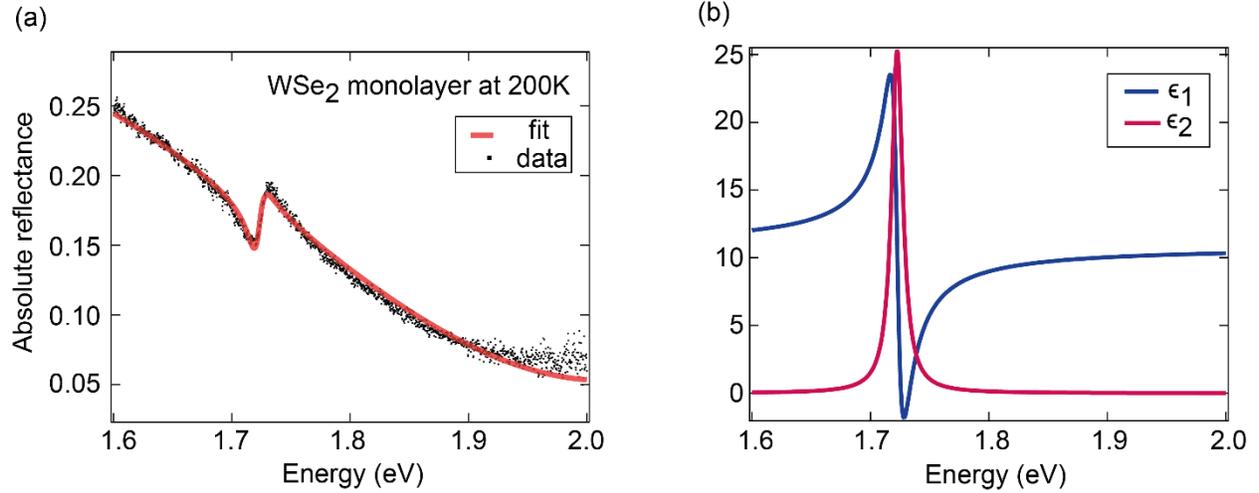

**FIG. S9.** Calibrating method by fitting dielectric function of monolayer WSe₂. (a) Absolute reflectance of WSe₂ monolayer on SiO₂/Si (310 nm) taken at 200K, along with the fit (red line) obtained using transfer matrix method. (b) The real and imaginary parts of the dielectric function obtained for WSe₂ monolayer by fitting.

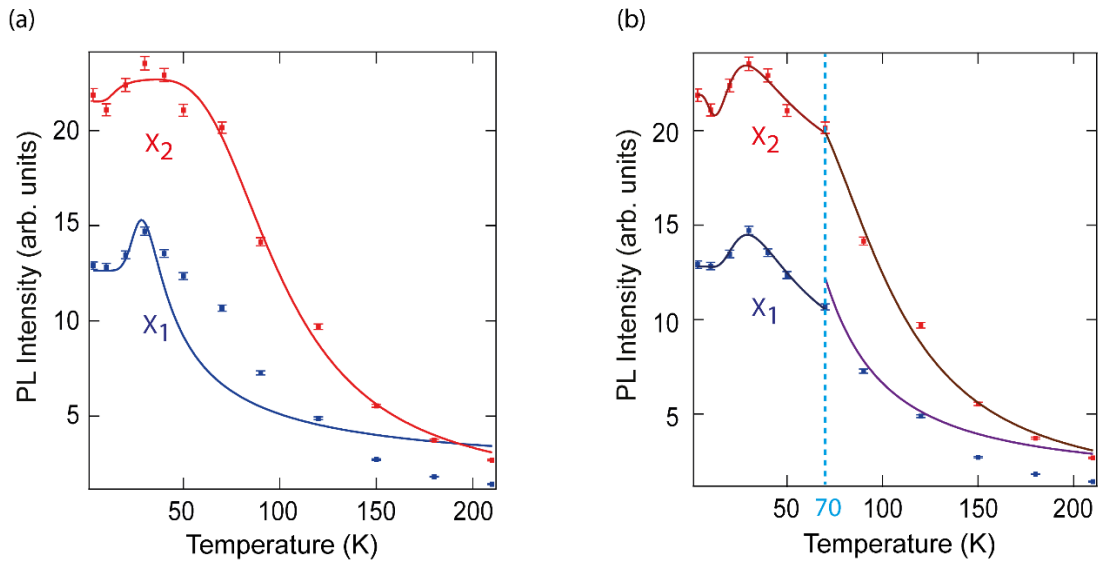

**FIG. S10.** Fitting obtained for temperature dependent PL intensity when using only one phonon scattering mode from each exciton. (a) Fit obtained when one phonon mode is used for the entire

temperature range. (b) Fit obtained when data for T≤70K and T≥70K is fitted separately with one phonon mode each, showing the need to consider two different phonon modes to account for scattering across all temperatures.

### 1. Fitting function to find the exciton orientation

In polarisation-resolved PL and differential reflectance measurement, the orientation of the excitonic dipole is determined by the fitting function obtained from Malus' law:

$$I = I_1 + I_0 \cos^2(\theta - \theta_d)$$

Where $I_1$ is a constant denoting the component of unpolarized light, and $I_0$ is also a constant, being the maximum value of the intensity. $\theta$ is the angle of the emission, for PL, and angle of input polarization for RC, w.r.t the b-axis. $\theta_d$ is the angle of orientation of the excitonic dipole moment w.r.t the b-axis.

### 2. Using Transfer Matrix Method to extract complex dielectric function of TMDCs on a substrate

For a multilayer system, the reflectivity can be conveniently calculated using the Transfer Matrix method, as discussed by Born and Wolf [1], and extended by Mcleod [2] to incorporate lossy materials with complex refractive indices.

For a homogeneous layer with complex refractive index *n* and width *a,* the transfer matrix is defined as:

$$T_a = \begin{pmatrix} cos\,ka & i/n\,sin\,ka \\ in\,sin\,ka & cos\,ka \end{pmatrix}$$

Where $k = (2\pi n/\lambda_0)$ is the value of the wavevector inside each layer. To obtain the reflectance *R* of the multilayer structure, the individual transfer matrices have to be multiplied in order, and then the resultant matrices elements are used as follows:

$$T = \prod_{i=m}^{1} T_i$$

$$r_s = \frac{n_{right}t_{11} + n_{left}n_{right}t_{12} - t_{21} - n_{left}t_{22}}{t_{21} - n_{left}t_{22} - n_{right}t_{11} + n_{left}n_{right}t_{12}}$$

$$R = |r_s|^2$$

In our case $n_{right}$ and $n_{left}$ are both taken to be the refractive index of air, unity. The equations can be easily modified for oblique incidence (k → kcosφ for TE polarization). Our reflectance measurement setup used an objective lens of NA 0.7 to focus light on and collect light from the sample, so some contribution from oblique incidence to reflectance was expected. However, while fitting, it was evident that normal reflectance was sufficient to model the obtained reflectance, since adding the effect of an angle spread made the fit worse. This agrees with the findings of a previous report [3] which found negligible contribution of higher angles up to NA 0.55. This also means that the contribution from TM polarized light incident on the sample can be ignored, since for smaller angles TE and TM reflectivity is practically the same.

To ascertain the exact thickness of $SiO_2$ in the substrate used, the absolute reflectance of the sample was first obtained using a different normal reflectance set up which used an Ag mirror as the

standard for reflectance. First the mirror kept at a certain height, and the reflected intensity $I_{mirror}^{ref}$ is recorded. Then the substrate is kept at the same height and the reflected intensity $I_{substrate}^{ref}$ recorded.

$$R_{substrate}^{abs} = \frac{I_{substrate}^{ref}}{I_{mirror}^{ref}} R_{mirror}^{abs}$$

$R_{mirror}^{abs}$ was obtained from the manufacturer specifications. To avoid introducing errors, we used the wavelength range starting about 400 nm for which the mirror has near-constant reflectivity. $R_{substrate}^{abs}$ was then fitted with theoretical reflectance of a Si/SiO$_2$ system, using literature values of refractive index for both materials [4,5] and the thickness of SiO$_2$ as a parameter – this allowed the determination of SiO$_2$ thickness across the substrate.

Armed with this, it is now possible to determine the complex refractive index of a thin film deposited on the substrate. We model the complex dielectric function of the material as a sum of Lorentz oscillators (for the excitonic resonances) and a background constant, $\epsilon(\omega) = \epsilon_b + \sum \frac{f_i}{\omega_{0i}^2 - \omega^2 - i\omega\gamma_i}$, with each oscillator contributing three parameters. $\epsilon_b, f_i, \omega_{0i}$ and $\gamma_i$ are the background dielectric constant, oscillator strength, resonance frequency and the linewidth of the $i^{\text{th}}$ oscillator, and the summation is over all exciton resonances. Oscillator frequencies used in this model are taken from the raw data. The parameters used for the fitting are $\epsilon_b$, $f_i$, $\gamma_i$ and the thickness of the ReS$_2$ layer which is then verified with our measured thickness value from the AFM (Fig. S1). All the fitted parameters are given in Tables S1-S4. Note that this functional form is automatically satisfies the Kramers- Kronig relations [6].The complex refractive index can be obtained from the dielectric constant through the relation:

$$\tilde{n}(\hbar\omega) = n(\hbar\omega) + ik(\hbar\omega) = \sqrt{\epsilon}$$

By choosing some initial guess values for the parameters, this is used to determine the reflectivity of the thin film/SiO₂/Si system $R_{sample}$, as well as the theoretical differential reflectivity:

$$R_{diff} = R_{sample}/R_{substrate}$$

This can be fitted with the experimentally obtained differential reflectivity, by varying the parameters using a non-linear fitting method, to give the correct complex refractive index. MATLAB was used for this purpose.

To check the efficacy of this method, we tested it by calculating the refractive index of WSe₂ on Si/SiO₂ at room temperature, the results of which are shown in Fig. S9. This agrees reasonably well with refractive index values obtained in previous studies [7,8].

Note that the reflectivity measurement of WSe₂ was done using unpolarized light. Modeling the reflectivity of ReS₂ obtained using unpolarized light led to unsatisfactory fitting results near the excitonic resonance energy range (as shown in Fig. S4) which tells us that the refractive index must be determined separately for each incident polarization angle, and not as an average over all angles, due to the inherent optical anisotropy of the material. There is no effective average refractive index for unpolarized light.

To obtain the absolute reflectivity of the sample (as shown in Fig. 2(c)), which was sometimes useful for quicker fitting, we used the theoretical reflectivity of the Si/SiO₂ substrate as the standard:

$$R_{sample}^{abs} = \frac{I_{sample}^{ref}}{I_{substrate}^{ref}} R_{substrate}^{abs}$$

The parameters used to obtain the fit in Fig. S4 and 2(c) are provided in Tables S1-4.

### 3. Rate equation model for temperature dependent photoluminescence intensity

The following rate equations are used to model phonon-scattering processes which are shown in Fig. 4(c) of the manuscript:

$$\frac{dx_{D1}}{dt} = g_{D1} - \frac{x_{D1}}{\tau_{D1}} - x_{D1}\gamma_1 e^{\left(-\frac{E_{01}}{k_B T}\right)}$$

$$\frac{dx_{D2}}{dt} = g_{D2} - \frac{x_{D2}}{\tau_{D2}} - x_{D2}\gamma_2 e^{\left(-\frac{E_{02}}{k_B T}\right)}$$

$$\frac{dx_1}{dt} = g_1 - \frac{x_1}{\tau_1} + x_{D1}\gamma_1 e^{\left(-\frac{E_{01}}{k_B T}\right)} - x_1\alpha_1 e^{\left(-\frac{E_{1n}}{k_B T}\right)} - x_1\alpha_2 e^{\left(-\frac{E_{1m}}{k_B T}\right)}$$

$$\frac{dx_2}{dt} = g_2 - \frac{x_2}{\tau_2} + x_{D2}\gamma_2 e^{\left(-\frac{E_{02}}{k_B T}\right)} - x_2\beta_1 e^{\left(-\frac{E_{2n}}{k_B T}\right)} - x_2\beta_2 e^{\left(-\frac{E_{2m}}{k_B T}\right)}$$

For CW excitation, the equilibrium condition is used (i.e., $\frac{dx_{D1}}{dt} = \frac{dx_{D2}}{dt} = \frac{dx_2}{dt} = \frac{dx_1}{dt} = 0$) to obtain the functional forms of PL intensities $I_1$ and $I_2$.

The number of parameters has been minimized while keeping in account all the physically important processes. The phonon scattering from $X_2$ to $X_1$ was ignored since it is negligible. The radiative lifetime is fixed to 10 ps based on values found in the literature. [9]

Considering two phonons for each exciton is necessary to capture the temperature variation in the lower and higher temperature regime. Excitons can scatter away from the light cone to a continuum of energy states. Therefore, phonons of all possible energies will participate in this. To effectively represent this scattering, we have picked only two phonon energies instead, one whose scattering will be dominant at low temperatures (<30K) and one at higher temperature regime.

If we use one phonon energy, the fitting will have significant departures from the experimental data in one of the temperature regimes, as shown in Fig. S10. Subfigure (a) is the fit obtained when only one phonon mode is used for the entire temperature range. Fig S10(b) also uses one phonon mode, but two temperature regimes (T≤70K and T≥70K) are fitted separately. The phonon energies obtained for low temp. is approximately 6 meV and for high temp is 33 meV, indicating the need to consider two energies to effectively account for scattering across all temperatures.

$X_3$ and $X_4$ were not explicitly considered in the model because it would increase the number of parameters further. The possible scattering from $X_3$ and $X_4$ down to $X_1$ and $X_2$ respectively is effectively accounted for in the carrier generation parameters $g_1$ and $g_2$. In addition, it was not possible to add $X_3$ and $X_4$'s intensity to the model since it is difficult to extract temperature dependent PL intensity for $X_3$ and $X_4$ since they are not resolvable beyond 150 K.

**Table S1.** WSe$_2$ model (200K)

Thickness: 6.49 Å, 310 nm SiO2
$\epsilon_b = 10.8$

| Oscillator no. | Origin | $\omega_0$ (eV) | $\omega_p$ (eV) | $\Gamma$ (meV) |
|---|---|---|---|---|
| 1 | A Exciton | 1.7226 eV | 0.707 eV | 11.40 meV |

**Table S2.** 11 nm ReS$_2$ model (for incident white light polarization at 220 degrees w.r.t. b-axis)

ReS$_2$ thickness = 11 nm ± 0.5 nm
SiO$_2$ thickness = 344 nm ± 1 nm
$\epsilon_b$ = 7.34

| Oscillator no. | Origin | $\omega_0$ (eV) | $\omega_p$ (eV) | $\Gamma$ (meV) |
|---|---|---|---|---|
| 1 | X1 | 1.546 | 0.296 | 6.44 |
| 2 | X2 | 1.578 | 0.23 | 5.28 |
| 3 | X3 | 1.559 | 0.126 | 10.60 |
| 4 | X4 | 1.601 | 0.251 | 27.43 |
| 5 | X1 low energy shoulder | 1.538 | 0.033 | 1.55 |
| 6 | X2 high energy shoulder | 1.583 | 0.135 | 6.16 |
| 7 | Rydberg | 1.637 | 0.102 | 10.15 |
| 8 | Rydberg | 1.664 | 0.248 | 48.85 |
| 9 | Drude term | 0 | 2.496 | 5265 |
| 10 | High energy peak | 2.232 | 4.028 | 364.11 |

**Table S3.** 11 nm $ReS_2$ model (for incident white light polarization at 270 degrees w.r.t. b-axis)

$ReS_2$ thickness = 11 nm ± 0.5 nm
$SiO_2$ thickness = 344 nm ± 1 nm
$\epsilon_b = 7.62$

| Oscillator no. | Origin | $\omega_0$ (eV) | $\omega_p$ (eV) | $\Gamma$ (meV) |
|---|---|---|---|---|
| 1 | X1 | 1.526 | 0.019 | 2.33 |
| 2 | X2 | 1.578 | 0.373 | 5.63 |
| 3 | X4 | 1.592 | 0.102 | 5.92 |
| 4 | Rydberg | 1.641 | 0.15 | 18.56 |
| 5 | Rydberg | 1.672 | 0.236 | 38.18 |
| 6 | Low energy background | 1.464 | 0.338 | 97.74 |
| 7 | High energy background | 1.746 | 0.705 | 202.88 |

**Table S4.** 11 nm $ReS_2$ model (for incident white light polarization at 350 degrees w.r.t. b-axis)

$ReS_2$ thickness = 11 nm ± 0.5 nm
$SiO_2$ thickness = 344 nm ± 1 nm
$\epsilon_b = 8.79$

| Oscillator no. | Origin | $\omega_0$ (eV) | $\omega_p$ (eV) | $\Gamma$ (meV) |
|---|---|---|---|---|
| 1 | X1 | 1.547 | 0.331 | 5.91 |
| 2 | X2 | 1.583 | 0.143 | 12.37 |
| 3 | X3 | 1.56 | 0.257 | 19.19 |
| 4 | Rydberg | 1.639 | 0.113 | 11.33 |
| 5 | Rydberg | 1.665 | 0.151 | 32.62 |
| 6 | Drude term | 0 | 2.903 | 2001.01 |
| 7 | High energy near band edge | 2.232 | 3.389 | 698.98 |

## Supplementary References



\end{document}